\documentclass[apj]{emulateapj}
\usepackage{natbib}
\usepackage[breaklinks,colorlinks,urlcolor=blue,citecolor=blue,linkcolor=blue]{hyperref}
\usepackage{amsmath}
\usepackage{graphicx}
\usepackage{verbatim}
\usepackage{epsfig}
\usepackage{xcolor}
\usepackage{commath}
\begin{document}

\shorttitle{Oxygen desorption from dust grains}
\shortauthors{He et al.}

\title{A new determination of the binding energy of atomic oxygen on dust grain surfaces: experimental results and simulations}
    \author{Jiao He\altaffilmark{1}, Jianming Shi, Tyler Hopkins and Gianfranco Vidali\altaffilmark{2}}
\affil{Physics Department, Syracuse University, Syracuse, NY 13244-1130, USA}

\author{Michael J. Kaufman}
\affil{Department of Physics and Astronomy, San Jos{\'e} State University, San Jose, CA 95192, USA}

\altaffiltext{1}{Current address: Department of Chemistry, University of Hawai'i at Manoa, Honolulu, HI 96822, USA}
%\email{gvidali@syr.edu}
\altaffiltext{2}{Corresponding Author: gvidali@syr.edu}

\begin{abstract}
The  energy to desorb atomic oxygen from an interstellar dust grain surface, $E_{\rm des}$, is an important controlling parameter in gas-grain models; its value impacts the temperature range over which oxygen resides on a dust grain. However, no prior measurement has been done  of the desorption energy. We report the first direct measurement of $E_{\rm des}$ for atomic oxygen from dust grain analogs. The values of $E_{\rm des}$ are $1660\pm 60$~K and $1850\pm 90$~K for porous amorphous water ice and for a bare amorphous silicate film, respectively, or about twice the value previously adopted in simulations of the chemical evolution of a cloud. We use the new values to study oxygen chemistry as a function of depth in a molecular cloud. For $n=10^4$ cm$^{-3}$ and $G_0$=10$^2$ ($G_0$=1 is the average local interstellar radiation field), the main result of the adoption of the higher oxygen binding energy  is that  H$_2$O can form on  grains at lower visual extinction $A_{\rm V}$, closer to the cloud surface.  A higher binding energy of O results in more formation of OH and H$_2$O on grains, which are subsequently desorbed by far-ultraviolet radiation, with consequences for gas-phase chemistry. For higher values of $n$ and $G_0$, the higher binding energy can lead to a large increase in the column of H$_2$O but a decrease in the column of O$_2$.
\end{abstract}

\section{Introduction}
Although oxygen is the third most abundant element in the universe, finding it has  proven to be challenging, especially in molecular form in dense clouds. Observations of molecular oxygen with the Submillimeter Wave Astronomy Satellite (SWAS) toward nearby clouds yielded no detection, while with the Odin satellite upper limits of 10$^{-7}$ (referred to H$_2$) were obtained for several cold dark clouds that were examined \citep{Pagani2003,Larsson2007}. \citet{Yildiz2013} put an upper limit of  6 $\times$ 10$^{-9}$ in a \emph{Herschel} HIFI search toward a low-mass protostar.  Detection with \emph{Herschel} HIFI of O$_2$ in Orion, via the rotational lines at 487, 774, and 1121 GHz, gave an estimated O$_2$ abundance of 10$^{-6}$ \citep{Goldsmith2011}; such a large value is attributed to the passage of a shock \citep{Chen2014}. Towards $\rho$ Oph A,  the estimated O$_2$ abundance is 5 $\times $10$^{-8}$ \citep{Larsson2007,Liseau2012}.  No detection  at 487 and 774 GHz toward the Orion Bar, an far-ultraviolet (FUV) illuminated region at the interface of H II region and a  dense molecular cloud,  was reported \citep{Melnick2012}. Overall these observations confirmed that molecular oxygen abundance in dense clouds is much less than predicted by gas-phase models, that is, 7 $\times$ 10$^{-5}$ for O$_2$ in a cold molecular cloud in steady state \citep[table 9 in ][]{Woodall2007}. \citet{Hincelin2011} looked at gas-grain oxygen chemistry and studied the sensitivity of rate coefficients on O$_2$ abundance. Motivated by the conflict between gas-phase models and observations, \citet{Hollenbach2009}  constructed a gas-grain model of the chemistry of molecular cloud surfaces and interiors. They found that the low O$_2$ abundances could be explained  by models in which O was largely atomic at cloud surfaces and locked primarily in water ice deeper into the cloud, with smaller contributions from CO, CO ice and refractory grain material. This model and others \citep[e.g.][]{Garrod2008,Hasegawa1993} highlight the importance of grain-surface reactions in understanding abundances in the interstellar medium. Depletion of oxygen in silicate grains is well constrained, as well as in CO and N-containing molecules \citep{Jenkins2009,Jenkins2014}. Oxygen is present in water, but  gas-phase water abundance is not enough to explain the missing oxygen. Water is also in ice-coated grains, but more water ice than it is observed in ice is necessary to account for all of the oxygen. Some water can be in hydrated silicates, on grains in low $A_{\rm V}$ regions before thick ices are formed, or on large ($>1\ \mu$m) grains; however, each of these proposed oxygen reservoirs has its own weaknesses \citep{Whittet2010}. It has  been suggested that oxygen is trapped in other types of grains, such as oxygen-bearing carbonaceous grains, similar to organics in cometary particles \citep{Whittet2010}.

On grains,  oxygen can react with hydrogen to yield OH and then H$_2$O via hydrogen addition reactions, or it can diffuse and form molecular oxygen and ozone (O is almost exclusively converted into OH/H$_2$O on the surface. O$_2$ is likely formed in the gas phase and eventually condenses on grain surfaces). From laboratory experiments, we know that O$_2$ leaves the surface of dust grain analogs at $\sim$ 30 K, but O$_3$ at a much higher temperature, $\sim$ 67 K \citep{He2014a}. O$_3$ can be hydrogenated on grains and form water \citep{Mokrane2009,Romanzin2011,He2014c}. In recent years several experiments were done on dust analogs in order to explore the routes to water formation via hydrogenation of O, O$_2$ and O$_3$ \citep[see references in][]{Vidali2013a}. How successful these routes are in forming water or hydroxyl products depends in large part on how successfully a dust grain keeps O on its surface. \citet{Tielens1982} derived the adsorption energy of O on grains, $E_{\rm b}$, by assuming that it is held by physisorption forces that depend on the atomic polarizability and the dielectric function of the solid \citep{Bruch2007}. Making an analogy with adsorption of gases on graphite for which an extensive literature exists \citep{Vidali1991,Bruch2007}, \citet{Tielens1982} derived $E_{\rm b}$=800 K. Eventually, this value was adopted in codes of simulations of gas-grain processes in the ISM \citep[e.g.][]{Stantcheva2002,Garrod2008}. One case where the models of \citet{Hollenbach2009} predicted high O$_2$ abundances ($\sim 10^{-5}$) was in regions with very high FUV fields, where desorption of O from grains would allow gas-phase production of O$_2$. Recognizing that observations were in conflict with this, they suggested the O binding energy was at least 1200 K. Observations of the Orion Bar by \citet{Melnick2012} set a stricter limit on the O$_2$ abundance in a high FUV region, leading these authors to propose that the binding energy was as high as 1600 K.  \citet{He2014a} used a rate equation model to analyze the data of O$_3$ formation on an amorphous silicate in the temperature range of 30--50~K and indirectly obtained the atomic oxygen adsorption energy, 1760 $\pm$ 230 K.

Here we present a direct measurement of the desorption energy $E_{\rm des}$ of atomic oxygen from porous amorphous water ice and from a bare amorphous silicate. (The desorption energy is the binding energy $E_{\rm b}$,  if there is no activation process, as we assume here in absence of evidence pointing otherwise). Then, we use the photodissociation region (PDR) model of \citet{Hollenbach2009}  to look at the consequences that  this revision has on the chemistry of the ISM by doing a simulation with the old and the new values of the desorption energy. We also look at the effect on ISM oxygen chemistry of using a revised value for OH desorption energy \citep{He2014c}. 

\section{Experimental}
Here we give a brief description of the experimental set-up and measuring methods; details can be found in \citet{Jing2013} and \citet{He2014a}. The apparatus consists of a triply differentially pumped atomic/molecular beam line connected to an ultra-high vacuum chamber. Molecular oxygen is dissociated in a  water-cooled Pyrex tubing placed in a radio frequency (RF) cavity. The O$_2$ gas flow is controlled by a mass flow controller. The beam enters the sample chamber through a 2.5 mm collimator. The narrow angular spread and the triple differential pumping of the beam line minimize the gas load to the sample chamber and the deposition of beam particles on the sample holder. The ultra high vacuum chamber is pumped by a turbomolecular pump, a cryopump and an ion pump. The background pressure is about 1 $\times$ 10$^{-10}$ torr when the system is  at room temperature and  $5\times10^{-11}$ torr when the sample holder column is cooled. When the beam enters the chamber, the pressure is of the order of 1-2 $\times$ 10$^{-10}$ torr. The detector is a Hiden Analytical HAL-5 quadrupole mass spectrometer (QMS) that is mounted on a doubly differentially pumped rotatable platform. Therefore, the detector can be swung in front of the beam to characterize the incoming beam and can be placed facing the sample during the deposition and the temperature programmed desorption (TPD). The detector itself  is placed in an enclosure that is differentially pumped. The aperture to the ionizer of the QMS is fitted with a Teflon cone with the apex facing the sample. The distance between the tip of the cone and the sample can be adjusted by repositioning the axis of the sample via an XYZ manipulator. The purpose of placing the cone is to maximize the flux coming from the sample and minimize the flux of particles coming from other parts of the sample holder. In this way, the signal/noise is enhanced. The procedures to calibrate the beam flux are described in \citep{He2015}. When the molecular oxygen beam is about 0.34 ML minute$^{-1}$ ($3.4\times 10^{14}\ \mathrm{cm}^{-2}\mathrm{minute}^{-1}$), the efficiency of dissociation is 42 \%. Thus the beam intensities of O and O$_2$ are 0.29 ML minute$^{-1}$ ($2.9\times10^{14}\ \mathrm{cm}^{-2}\mathrm{minute}^{-1}$) and 0.20~ML~minute$^{-1}$ ($2.0\times10^{14}\ \mathrm{cm}^{-2}\mathrm{minute}^{-1}$), respectively. 

The preparation and characterization procedures of the 1 $\mu$m thick amorphous silicate film have been described before \citep{Jing2013}. The sample was mounted on a sample holder that was cooled with liquid helium. Prior to a set of experimental runs the sample was taken to 400 K for cleaning. We found that this procedure yields the same results as by sputtering the sample with Ar ions \citep{Jing2013}. The sample temperature was measured by a calibrated silicon diode placed behind the sample; a thin silver foil assures a good thermal contact between sensor and sample. There is a thermal switch between the sample holder and the liquid helium stage that allows the heating of the sample without warming up other parts of the apparatus. This assured a lower background gas pressure during the experiments. 

During the atomic oxygen (O) exposure, because the dissociation rate is not unity, there is always O$_2$ mixed with O, so ozone formation via O+O$_2$ $\rightarrow$ O$_3$ is likely. When O desorption is seen in the TPD, it could come from the bare silicate surface and/or  from ozone that was formed on the silicate as explained above. To find out which one is true, the silicate surface was pre-covered with different coverage of ozone before depositing atomic oxygen. If O desorbs from the silicate surface, then as the pre-covered ozone approaches one monolayer, the amount of O desorption should decrease to zero. In the other case, if O desorbs from an ozone patch, then as the pre-covered ozone amount increases from zero to one monolayer, the amount of O desorption should increase linearly. The correlation between the coverage of ozone and the amount of atomic oxygen direct desorption in the TPD can suggest whether O is desorbing from the silicate surface or from ozone. It will be shown later that the former is true. The presence of ozone on the silicate surface  actually helps to decrease the formation of O$_2$ and O$_3$ via, respectively, the O+O and  the O+O$_2$ reactions, since O and O$_2$ do not react with O$_3$.  Ozone functions as an obstacle in the diffusion of O and O$_2$ on the surface. The procedure to prepare  the sample with a layer of ozone was described in \citet{He2014a}. Briefly,  the silicate surface was exposed to the dissociated oxygen beam and ozone was formed by O+O$_2$ $\rightarrow$ O$_3$; then the surface was annealed at 50 K to desorb unreacted O$_2$. During the exposure stage of ozone preparation, the beam flux was set to a higher value than that for the O deposition, so that the dissociation rate was lower. The beam intensity of O$_2$ was higher than O so that almost all the O was reacted.

In dense clouds, dust grains are usually covered by ice mantles, with water as the main component. We also carried out measurements on porous amorphous water ice. The ice substrate was prepared with background deposition through a microcapillary doser. The thickness was estimated from the background pressure and the duration of the deposition. A 500 ML (with an uncertainty of 30 \%) porous amorphous water ice film was deposited at 70 K on top of a 500 ML single crystalline water ice. The ice film was cooled down to lower temperatures for O/O$_2$ beam irradiation and TPD. During the following TPD experiments, the porous ice film was maintained below 70 K to prevent pore collapse in the ice films (see the temperature ramp in Figure \ref{porous}).

\subsection{Results and analysis}
\begin{figure}[t]
\epsscale{1}
\plotone{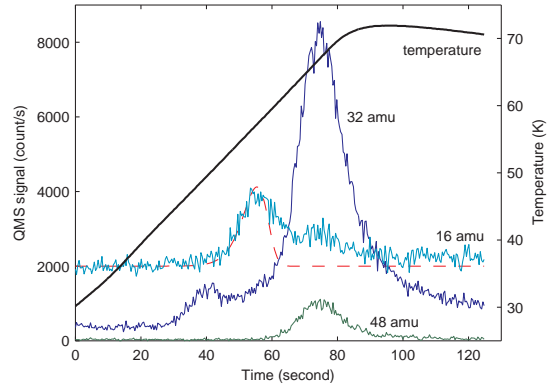}
\caption{TPD traces of mass 16, 32, and 48 amu after depositing 480 s of O/O$_2$ at 48 K on porous amorphous water ice. The heating ramp is 0.17 K~s$^{-1}$. The dashed line shows a fitting of the mass 16 amu peak using $E_{\rm{des}}(\rm O)= 1660$~K (see text).}
\label{porous}
\end{figure}

Figure \ref{porous} shows TPD spectra of oxygen from a porous amorphous water ice film. We deposited 480 s of O/O$_2$ on porous amorphous water ice at 48 K. After the exposure, the surface was cooled down to about 20 K and then heated up at a linear ramp rate of 0.167 K~s$^{-1}$ to 70 K to desorb the atoms and molecules on the surface. During the TPD, the desorption rate was recorded by the QMS facing the sample. Ozone was formed on the ice and came off at about 67 K. Because ozone is fragile against electron impact ionization, a good fraction breaks up into O$_2^+$ and O$^+$, as it can be seen from the peaks of mass 16 amu and 32 amu at about 67 K. In the mass 16 amu signal, there is a peak at 57 K, which is due to O direct desorption. As the sample was heated, O diffuses and forms O$_2$, and O$_2$ leaves the surface upon formation, because of the short residence time of O$_2$. The mass 32 amu peak centered at about 50 K is due to O$_2$ formation via O diffusion. As O is gradually used up, the O$_2$ formation rate decreases, and the rate of O direct desorption increases. The TPD peak area of O direct desorption corresponds to about 3\% of a layer. 

\begin{figure}[ht]
\epsscale{1}
\plotone{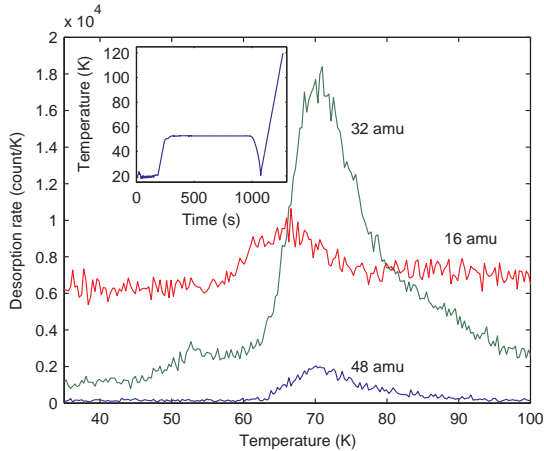}
\caption{ TPD traces of 16, 32, and 48 amu  after depositing  480 s of O/O$_2$ at 48 K on an amorphous silicate that has been pre-coated with 0.2 ML of ozone.The heating ramp is 0.5 K~s$^{-1}$. The inset shows the temperature ramp. }
\label{amorphous_2+8}
\end{figure}

\begin{figure*}[ht]
\vbox{
\epsscale{0.53}
\plotone{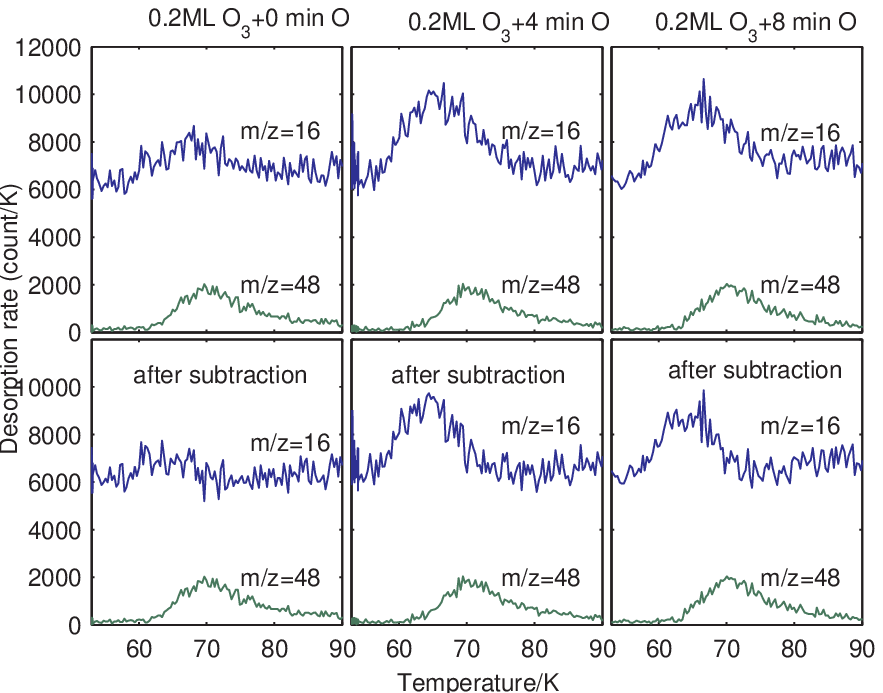}
\epsscale{0.53}
\plotone{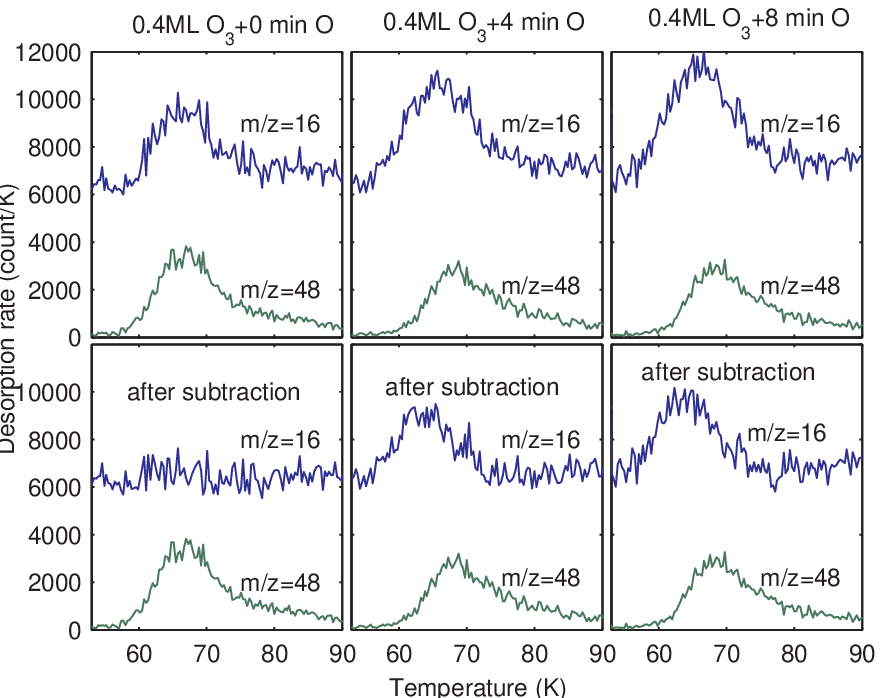}
\epsscale{0.53}
\plotone{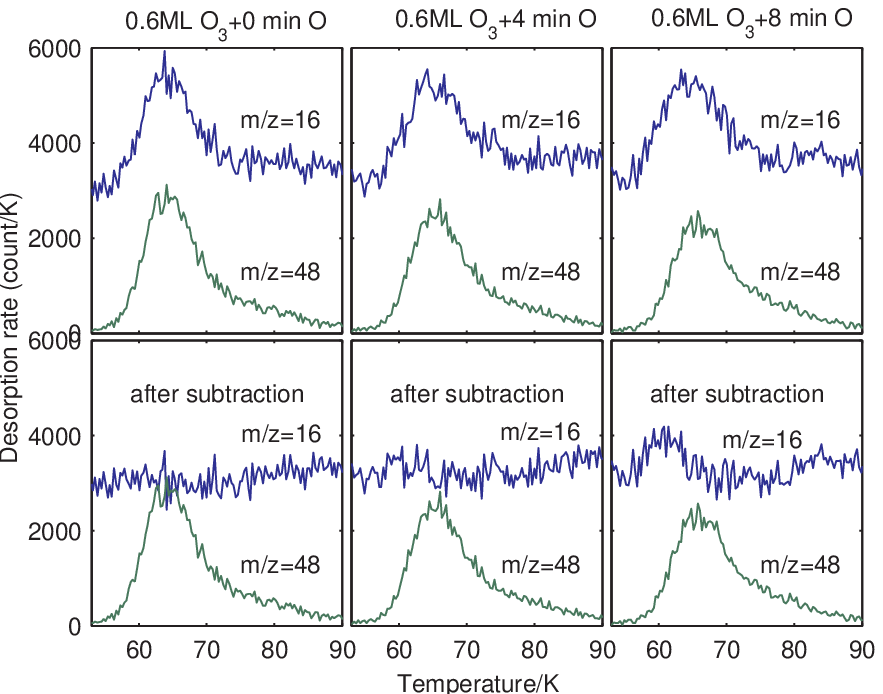}
\epsscale{0.53}
\plotone{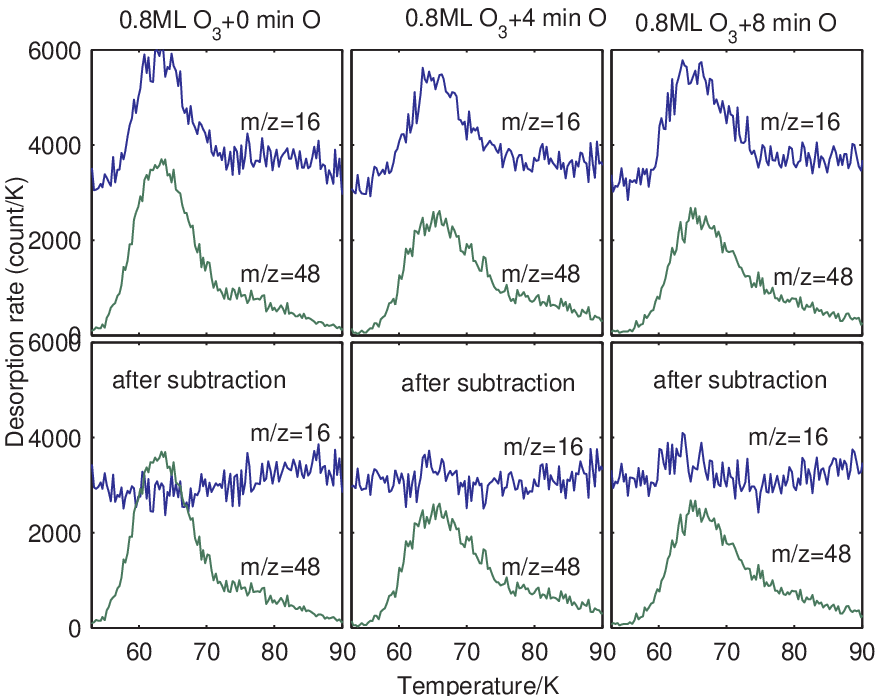}
}
\caption{ TPD traces of mass 16 amu (blue) and mass 48 amu (green) after deposition of 0, 240, and 480 s of O/O$_2$ on 0.2, 0.4, 0.6, and 0.8 ML of O$_3$ pre-coated amorphous silicate. The heating ramp is 0.5 K~s$^{-1}$. The top row of each panel has the original TPD traces while in the bottom row the contribution of O$_3^+$ fragmentation to the signal of mass 16 amu has been subtracted.}
\label{O3+O}
\end{figure*}

Similar experiments were performed on the surface of an amorphous silicate. Compared to porous amorphous water ice surface, the amorphous silicate surface is flatter. The mobilities of O and O$_2$ on it are higher, and the formation of O$_3$ and O$_2$ are also faster. To hinder the formation of O$_2$ and O$_3$, and leave more atomic oxygen on the surface available for direct desorption, we pre-coated the surface with a fraction of a layer of O$_3$.   O$_3$ does not react with atomic oxygen or molecular oxygen; O$_3$ is used to hinder the diffusion rate of atomic and molecular oxygen. To find out whether atomic oxygen is desorbing from bare amorphous silicate surface or ozone patch, measurements were performed to check the amount of O direct desorption from silicate surface covered with 0.2, 0.4, 0.6, and 0.8 ML of ozone. Figure \ref{amorphous_2+8} shows the TPD spectra after depositing 480 s of O/O$_2$ on a 0.2 ML O$_3$ pre-coated amorphous silicate surface. In the figure, the two contributions to the mass 16 amu peak, i.e., O direct desorption and O$_3^+$ fragmentation, are merged into a single broad peak. To find out the peak position and peak area of O direct desorption, the O$_3^+$ contribution needs to be subtracted out. The fragmentation pattern of O$_3^+$ is measured by O$_3$ TPD experiments and not shown here. The raw TPD data and that after subtracting O$_3^+$ contributions are shown in Figure \ref{O3+O}. In each panel of this figure, the first row has the raw TPD data, while in the bottom row the contributions of ozone fragmentation to mass 16 amu have been subtracted to show the amount of direct O desorption. After this subtraction, the O direct desorption peak is centered at about 64 K and is clearly different from the ozone desorption peak. We also found that as the coverage of the pre-coating ozone layer increases from 0.2 ML to 0.4 ML, the amount of O direct desorption increases. This is because the increased ozone amount helps to lower the O diffusion rate.  However, when the pre-coating ozone is more than 0.6 ML, there is almost no O direct desorption peak. This indicates that the observed O direct desorption is from bare silicate surface instead of ozone ice; otherwise, the O direct desorption amount should be positively correlated with the amount of ozone pre-coating. We also checked that the desorption energy of O atoms is not influenced by the presence of ozone at low coverage. In a work by \citet{Bennett2005}, the formation of a O$_3$--O complex was deduced from IR data in an experiment of bombardment of an O$_2$ ice at 10K by 5keV electrons. In our experiment we find that the mass 16 amu peak shape and temperature, see Figure 3, do not change as a function of O$_3$ pre-coverage (at low coverage) suggesting that there is no effect of a possible O$_3$--O interaction on the desorption on O. Furthermore, there is no reaction of  O$_3$ with  O because of an activation energy barrier \citep{Atkinson2004}.

\begin{figure}[ht]
\epsscale{1}
\plotone{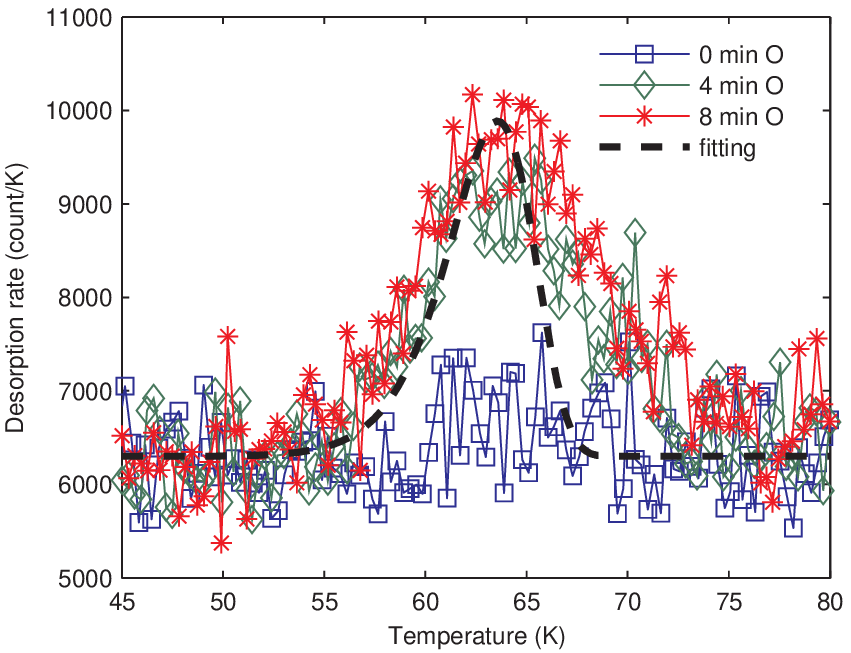}
\caption{Fitting of the mass 16 amu TPD traces after subtraction in the top-right panel (4 min O$_3$ + 0, 4, 8 min O) of Figure \ref{O3+O} using $E_{\rm des}=1850$~K.}
\label{4minO3+O_fit}
\end{figure}

The desorption energy of atomic oxygen from different surfaces can be calculated using an Arrhenius-type expression:
\begin{equation}
 \frac{\dif N(t)}{\dif t} = -\nu N(t) \exp(\frac{E_{\rm des}}{k_\mathrm BT(t)}) 
\end{equation}
where $N(t)$ is the number of molecules on the surface at time $t$; $\nu$ is a pre-exponential factor which is standardly taken to be $10^{12}\ \mathrm s^{-1}$; $E_{\rm des}$ is the desorption energy; $k_\mathrm B$ is the Boltzmann constant; $T(t)$ is the surface temperature. In Figure \ref{porous} the dashed line shows the fitting of mass 16 amu trace using $E_{\rm des}=1660$~K. In Figure \ref{4minO3+O_fit}, the mass 16 amu traces in the top-right panel of Figure \ref{O3+O} after subtraction of O$_3$ contribution are plotted together with a fitting using $E_{\rm des}=1850$~K. The energy values are shown in Table \ref{tab:O_energies}. The error bars of the desorption peak temperature are determined by the uncertainty in locating the peak positions. The width and shape of the TPD peak are not taken into account. Therefore, the error bar of desorption energy does not represent the width of distribution; instead, it  represents only the uncertainty in finding the peak of the TPD trace. 
\begin{table}[ht]
\caption{Desorption energy of atomic oxygen from porous amorphous water ice and from an amorphous silicate surface.}
\begin{center}
\begin{tabular}{ccc}
\hline
Surface & Desorption Peak Temperature & $E_{\rm{des}}(\rm O)$\\
\hline
Porous water ice & 56$\pm 2$ K & 1660$\pm 60$~K\\
Amorphous silicate  & 64$\pm 3$ K & 1850$\pm 90$~K\\
\hline
\end{tabular}
\end{center}
\label{tab:O_energies}
\end{table}

\section{Simulations and Astrophysical Implications}

In order to explore the implications of a higher oxygen desorption energy, we have used the PDR model of \citet{Hollenbach2009}  which computes the chemical and thermal structure of a molecular cloud edge exposed to FUV radiation from either the interstellar radiation field or from nearby hot stars. The important input parameters are the gas density, $n_{\rm H}$, the strength of the FUV radiation field, $G_0$ (where a value $G_0=1$ corresponds to the FUV radiation field in the diffuse interstellar medium), the abundances of important atoms, and the properties of interstellar dust grains. With these inputs, the model self-consistently computes the gas and grain temperatures, and the abundances of more than 60 species in the gas phase and in ice mantles on grain surfaces, starting from the cloud surface and extending in to extinctions $A_ \mathrm V>10$, deep within the shielded inner layers of a molecular cloud. For the comparisons below, we use the same chemical abundances, grain properties and rate coefficients as in \citet{Hollenbach2009} with the exception of the binding energies of O and OH on grain surfaces. For the binding energy of O, we adopt a characteristic value from this work, $E_{\rm b}({\rm O})=1800$~K;  for the binding energy of OH, we adopt the highest value from \citet{He2014c}, $E_{\rm b}({\rm OH})=4800$~K, though the precise value is not important so long as $E_{\rm b}({\rm OH})>E_{\rm b}({\rm O})$, as discussed below.  Here we assume that $E_{\rm des}=E_{\rm b}$, i.e., that the binding (adsorption) and desorption energies are the same, which is true if the processes are not activated.

In the models, the abundances of species on dust grains are controlled by competition between adsorption onto grain surfaces, chemical processes which occur on the grain surfaces, and desorption processes---including photodesorption by FUV photons, thermal desorption from warm grains, and cosmic ray desorption. The chemistry of oxygen-bearing molecules was shown to be driven by grain-surface reactions. The chemistry of water, in particular, begins with the adsorption of atomic oxygen onto a grain surface. If the grain encounters a hydrogen atom before the oxygen atom is desorbed from the surface, then OH ice is formed.  In the model, it is assumed that H atoms are highly mobile on the grain surface can easily find O in ice. If the OH is at least as tightly bound to the surface as  the O, then OH will also abstract an H, forming water ice on the grain.  This grain-surface ice is then the source, via photodissociation, of gas phase water near the cloud surface.  The main result of the increased O binding energy described in this work is that O will stick to grains that are warmer, so that the grains can catalyze water formation in higher FUV fields and closer to the cloud surface. 

Following the discussion in \S2.5 of \citet{Hollenbach2009}, we find that the new value of the O binding energy should allow formation of water on grains for grain temperatures as high as $\sim$ 50~K. In the models, the grain temperature in the unshielded cloud surface is $T_{\rm gr}\sim 12\ G^{0.2}$~K. Thus grain surface formation of water should proceed efficiently near the cloud surface for FUV fields as high as $G_0\sim 10^3$, and may also be efficient in regions with higher FUV fields deeper into the clouds where the grain temperature has dropped to $\sim 50$~K. 

\begin{figure}[ht]
\epsscale{0.9}
\plotone{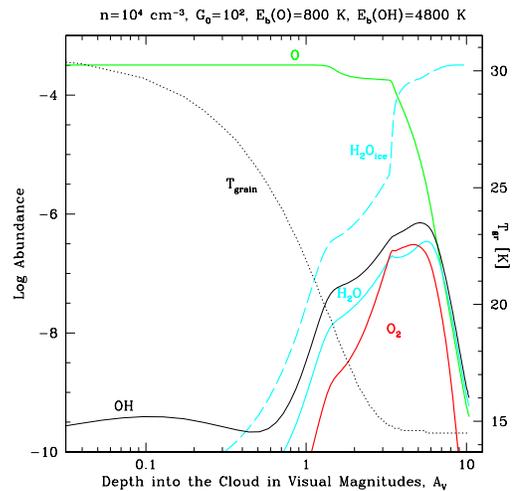}
\caption{ Photodissociation region model for $n=10^4\,{\rm cm^{-3}}$ and $G_0=10^2$. with O binding energy of 800~K. Left axis shows the abundances of O-bearing species as a function of cloud depth, while the right axis indicates the grain temperature. The dashed line shows the abundance of water ice on grains.}
\label{modelfig5a}
\end{figure}

\begin{figure}[ht]
\epsscale{0.9}
\plotone{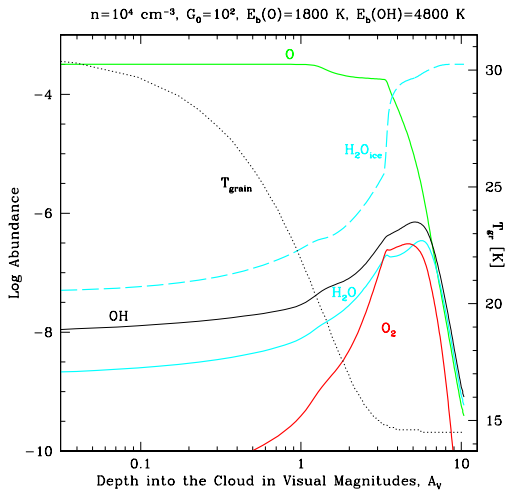}
\caption{ Same as Fig. \ref{modelfig5a} but with an O binding energy of 1800~K.}
\label{modelfig5b}
\end{figure}

To illustrate the effects of the increased binding energies, we have computed models with $E_{\rm b}({\rm O}) = 800$~K and with $E_{\rm b}({\rm O})=1800$~K for two different combinations of $n$ and $G_0.$ In the first case, we compute the results for models with $n=10^4\,{\rm cm^{-3}}$ and $G_0=10^2$, conditions appropriate for a molecular cloud interface near the outer edge of an \ion{H}{2} region. In Figure \ref{modelfig5a}, we show the results for $E_{\rm b}({\rm O}) = 800$~K and in Figure \ref{modelfig5b}, we show results for $E_{\rm b}({\rm O}) = 1800$~K. The effects may be seen in the outer edge of the cloud, where the grain temperature is above 30~K. In the case of the old binding energy, O does not stick to grains and begin to form water ice until the grain temperature has dropped below $\sim$25~K, at $A_{\rm V}\sim 0.5$. For the higher O binding energy, O can stick to grains all the way out to the cloud surface, where the grain temperature is $\sim 30$~K. However, the strong FUV field is able to photodesorb water ice from the grains at a rate high enough to prevent a monolayer of ice from forming. Thus, although the water ice and gas phase water abundances in the outer regions of the cloud are several orders of magnitude higher than if the O binding energy were lower, the effect on the total column densities of water and other O-bearing species is negligible. This model simply illustrates the effect of the binding energy on the ability of O to reside on grains. 

\begin{figure}[ht]
\epsscale{0.9}
\plotone{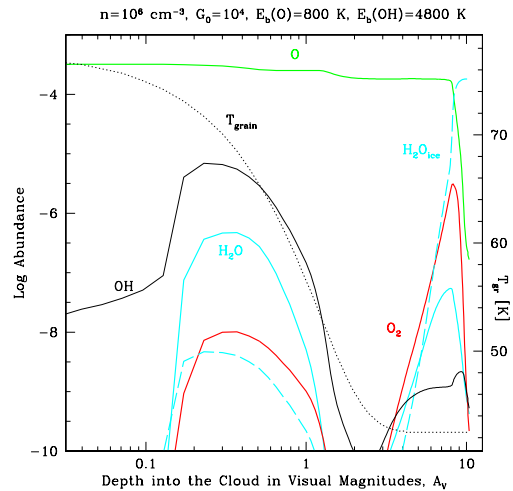}
\caption{ Photodissociation region model for $n=10^6\,{\rm cm^{-3}}$ and $G_0=10^4$ with O binding energy of 800~K. Left axis shows the abundances of O-bearing species as a function of cloud depth, while the right axis indicates the grain temperature. The dashed line shows the abundance of water ice on grains.}
\label{modelfig6a}
\end{figure}

\begin{figure}[ht]
\epsscale{0.9}
\plotone{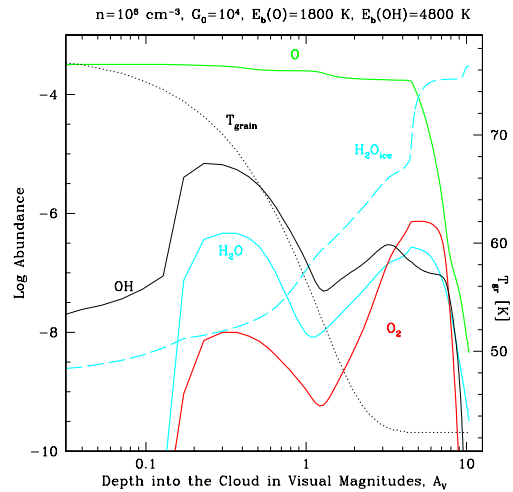}
\caption{ Same as Fig. \ref{modelfig6a} but with O binding energy of 1800~K.}
\label{modelfig6b}
\end{figure}

A more extreme case can be seen in Figs. \ref{modelfig6a} and \ref{modelfig6b}. Here we make the same comparison as in Figs. \ref{modelfig5a} and \ref{modelfig5b} but with $n=10^6\,{\rm cm^{-3}}$ and $G_0=10^4$, about the most extreme values expected in a dense star forming region like Orion or a dense nuclear starburst. Here there are two possible routes for water formation. In the outer layers of the cloud, the {\it gas} temperature can be several hundred degrees, high enough to drive the neutral--neutral reactions O + H$_2$ $\rightarrow$ OH + H and  OH + H$_2$ $\rightarrow$ H$_2$O + H, forming water directly in the gas phase. In addition, O$_2$ can form from O + OH. These reactions are responsible for the broad water peak in the two figures at $A_{\rm V}\sim 0.3.$  Beyond this depth, the gas temperature is too low to drive these reactions, and water formation must depend on the grain surface mechanism. In the model with low binding energy, O stays off the grains beyond $A_{\rm V}$ $\sim$1 and is driven into O$_2$ which reaches an abundance $\sim 10^{-5}$. The fact that this was not observed is precisely what led \citet{Hollenbach2009} to suggest a higher binding energy. The effects of the higher binding energy on H$_2$O and O$_2$ abundances can be seen in Fig. \ref{modelfig6b}. Once the grain temperature drops below 60~K, the model with the high value of the O binding energy can begin to form water ice on grain surfaces. The result is an increase in the column density over the lower binding energy case of a factor $\sim 4$, with a corresponding increase the in the intensity of optically thick but effectively thin water emission lines. The effect on the column of molecular oxygen, however, is in the other sense; with a higher value of the O binding energy, O is locked up in water ice and can not form O$_2$ by the O + OH reaction as it would if O were still in the gas phase. The result is a {\it lower} gas-phase column of O$_2$ than had been predicted for high $G_0$ cases in \citet{Hollenbach2009}. In another paper, we will more extensively explore the effects of the new binding energies. 

\section{Summary}
Oxygen is the third most abundant element, and significant observational and theoretical work has been dedicated to understanding its distribution in the interstellar medium. In this work, we explored the possibility that oxygen might reside on grains over a wider temperature range than previously accepted. In the first measurement of oxygen binding to dust grain analogs, we find that the values of the binding energy (1660 and 1850~K for O on, respectively,  porous amorphous water ice and a bare amorphous silicate) are more than twice the previous estimate \citep[800~K,][]{Tielens1982}. Our values are near the ones that were suggested in order to reconcile observations with abundance constraints \citep[1200~K and 1600~K proposed by, respectively,][]{Hollenbach2009,Melnick2012}. Since the estimated value (800~K) has been widely used in simulations of the chemical evolution of ISM environments, we performed simulations with the old and new values using as a test the same ISM environment as in \citet{Hollenbach2009} and \citet{Melnick2012}, i.e., a molecular cloud edge exposed to FUV. For a moderate illumination ($G_0$=10$^2$), the abundances of H$_2$O$_{\rm ice}$ and H$_2$O$_{\rm gas}$ increase by orders of magnitude, although it leaves the column density marginally affected. More interesting is the case of higher FUV illumination ($G_0=10^4$), a case expected in a dense star forming region as in Orion. The higher gas temperature allows neutral--neutral reactions to form OH and H$_2$O. At $A_{\rm V}\sim 1$, the gas temperature has dropped enough that water ice formation occurs aplenty on surfaces of grains. While the column density of water increases, the column density of gaseous O$_2$ decreases because more oxygen is locked up in grains. 

\section{Acknowledgments}
This work is supported by the NSF Astronomy and Astrophysics Division (Grant No.1311958 to GV) and by NASA support for US research with the \emph{Herschel Space Observatory} (RSA No. 1427170 to MJK). We thank Dr. J. Brucato of the Astrophysical Observatory of Arcetri (Italy) for providing the sample. We thank David Hollenbach for helpful suggestions.

\end{document}